# F-theory amplitudes

Warren Siegel

*CNYITP*
*Stony Brook University*

and

Yu-Ping Wang

*Dept. of Physics and Astronomy*
*Stony Brook University*

November 19, 2020

### Abstract

We propose 4-point S-matrices for three-dimensional F-theory. We will use the twistor formalism to facilitate constructing the amplitude. We write the amplitude in a way such that the F-symmetry (U-duality symmetry) is manifest. The amplitude can be schematically written as $A_4 = w^4/stu$, where $w$ is an analog of the linearized Weyl tensor in F-theory, and $w^4$ is a shorthand for the sum of various contractions that can happen between the Weyl tensors. The gauge invariance is actually non-trivial since $w$ is in general not gauge invariant. With the help of the twistor formalism, one can verify that this formula is indeed gauge invariant. The amplitude also reduces to the ordinary 4-graviton amplitude under the reduction to M-theory (which is just 4D supergravity).

---

This paper is compiled in LuaTeX. The source code can be found here.

# Contents

# 1 Introduction

F-theory is a theory that tries to make S, T and U-duality of string theory manifest [1–6]. More precisely, when compactifying M-theory to $\mathbb{R}^{11-(D+1)} \times T^{D+1}$, inside the $D+1$-dimensional torus the winding modes and the momentum (which is now also discrete) enjoy a discrete $E_{D+1}(\mathbb{Z})$ U-duality symmetry. It includes both the $O(D, D, \mathbb{Z})$ T-duality symmetry on the type II strings, and $GL(D+1, \mathbb{Z})$ S-duality symmetry on the original M-theory [23].

It is proposed that instead of treating the duality symmetry as a discrete symmetry on the quantum level, one makes the $E_{D+1}$ symmetry continuous on the classical level. This is achieved by adding extra coordinates in the internal space which correspond to the "momentum" and "winding mode" in the compactified space. Thus, instead of a $D+1$ dimensional torus, we have something usually referred to as the extended space, on which a theory with manifest $E_{D+1}$ symmetry is given [10].

It is found that the extended space can have an unusual geometric structure. It can be endowed with a vielbein in the coset space $G/H$ (see details below), and a generalized Lie derivative. The generalized Lie derivative generates coordinate transformations on the extended space. Since the generalized Lie derivative doesn't obey the Jacobi identity, the coordinate transformation is not closed. We have to impose section conditions to make it close. This space with a new Lie derivative and vielbein form the basis for the Generalized geometry. Solving the section condition we can break the $E_{D+1}$ symmetry to get the original M-theory [11–14].

While the Lagrangian for general $D$ is not found, the general current algebra with $E_{D+1}$ symmetry manifest is found (i.e. The Hamiltonian formalism) [7, 9].

For this paper, we are interested in finding the amplitude of F-theory, written in a way that the $E_{D+1}$ symmetry is manifest. More specifically, we are going to write down the 4-pt amplitude with manifest $E_4$ symmetry ($D = 3$) of the massless bosonic sector. To find this amplitude, several ingredients need to be in place. The most important of them is the twistor formalism for F-theory [15]. It will help us to both write the momentum in a way that is on-shell but also preserve the original off-shell symmetry.

In section 2, we will introduce the ingredients that are needed to construct the 4-pt amplitude. In section 2.1, we will give a quick introduction to the coset construction of the metric, and how the generalized coordinate transformation acts on the metric. The section condition will restrict our momentum; solving the constraint will give us back the original $D+1$-dimensional vector. On the other hand, one can reduce F-symmetry in a different way to T-symmetry. (In the current algebra construction, solving the Gauss constraint will reduce F-theory to T-theory.) If we break the section condition to T-theory, we will find it becomes the usual section condition for T-theory.

In 2.2, we will discuss how we can write the 4-pt amplitude of ordinary gravity in a way that gauge invariance is manifest (without choosing the polarization vectors explicitly). Namely, $A_4 = w^4/stu$, where $w$ is the on-shell Weyl tensor. For F-theory, we will try to write the amplitude in a similar way. Therefore, we need to find the Weyl tensor for the F-theory first. Fortunately, for the $D = 3$ case, the indices are simple enough such that the form $\partial^2 h$ has only two possible contractions. By going on-shell and writing everything in terms of the twistors, one of them will become 0. Unfortunately, the "Weyl tensor" we get is not gauge invariant. This is somewhat expected since it is speculated in other literature that there is no tensor in generalized geometry that corresponds to the Riemann tensor in general relativity [2, 11]. As we will see in section 6, while the individual Weyl tensor is not gauge invariant, the scalar contractions of any number of Weyl tensors are gauge invariant. We can find a unique contraction of $w^4$ that will reduce to the kinetic factor of the ordinary gravity amplitude when reduced to M-theory; this is the correct F-theory amplitude we are looking for.

In section 3, we will develop the twistor formalism of F-theory. We will focus on the case of $D = 3$. For 3D F-theory, the momentum can be represented as a bi-spinor $P_{(\alpha\beta)}$, where $\alpha, \beta$ are $Sp(4, \mathbb{R})$ indices. If $P_{(\alpha\beta)}$ is on shell, we can factor $P_{(\alpha\beta)}$ into $\lambda_{(\alpha}\bar\lambda_{\beta)}$, where $\lambda, \bar\lambda$ are in the coset space $H_D/H_D$, where $H_D$ is the local isometry group of F-theory (the analogue of the local Lorentz group for general relativity), and $H_D$ is the little group. (See 4 for more explanation.) For $D = 3$, $H_D = Sp(4, \mathbb{R})$ and $H_D = U(1)$. $\lambda_\alpha, \bar\lambda_\beta$ are analogous to twistors for ordinary 4D momentum. (We should also include the dual twistors $\partial/\partial\lambda, \partial/\partial\bar\lambda$, but we won't use them here.) The advantage of using twistors is that it makes the symmetry $H_D$ manifest, thus making it easier to construct quantities that are gauge invariant. The twistor formalism plays a crucial role in determining the gauge invariance of the 4-pt amplitude we construct. Since the momentum in F-theory obeys the section condition, we can write the section condition in terms of the twistors (which we will call the symplectic constraint). Solving the symplectic constraint, we reduce the F-theory twistor into the M-theory twistors, which are just the ordinary 4D momentum twistors.

In section 4, we find the metric $h$ for $D = 3, 4$, and break them down into $H_D$ spinor indices. We are interested in the perturbation under flat background, so flat and curved indices are indistinguishable. The (perturbed) on-shell metric can be written in terms of twistors of the momentum and some reference twistor. The choice of different reference twistors reflects a gauge transformation on $h$. If we plug the relation between $h$ and twistors into the Weyl tensor, we can represent $w$ in terms of the twistors, too.

In section 5, we break down the twistors, metrics, and Weyl tensors into M-theory and T-theory; we can see that for the M-theory case, only the component that is non-zero is the component containing 4D gravity, which is as expected.

# 2 Reviews

## 2.1 Cosets

F-theory, as the theory with manifest S, T, U-duality, can be defined using coset construction. The F-theory cosets are those of the scalars of maximal supergravity with $D$ internal dimensions, from compactification and duality transformations of $D$ dimensions of all the massless, bosonic fields of the 10D Type II superstring (or $D+1$ dimensions of 11D supergravity). One can assemble the scalars into the coset space $G/H$. $G$ is the Lie group $E_{D+1}$ and $H \equiv H_D$ a subgroup of $G$. (For the exact group of $H$ in various $D$ see below.) The vielbein is written as $E_A{}^M$, where the global group $G$ acts on the curved index $M$ while $H$ acts locally on the flat index $A$. In the rest of the paper, we will refer to indices $A, B \cdots$ as flat indices of the coset, which are in various representations that depend on $D$. They are not necessarily the fundamental representations; see Appendix A.

F-theory can break down into M and T-theory by solving certain section conditions or worldvolume constraints. The pattern of symmetry breaking due to sectioning and worldvolume constraints is indicated by the "diamond" (where arrows indicate subgroups) [2]:

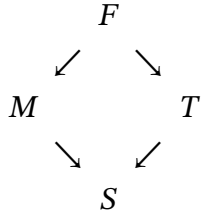

The cosets that incorporate the (generalized) metric (vielbiens) of these theories are defined as follows:

- The $D$-dimensional S-theory coset space $G/H$ is just gravity

$$GL(D)/O(D-1,1) \quad (S)$$

- The M-theory coset is gravity in 1 higher (space) dimension

$$GL(D+1)/O(D,1) \quad (M)$$

- $D$-dimensional T-theory has the coset

$$O(D,D)/O(D-1,1)^2 \quad (T)$$

of the scalars of $D$ internal (compactified) dimensions, resulting from compactification of the metric and 2-form of the orientable, closed bosonic string.

- $D$-dimensional $F$-theory has the coset

$$E_{D+1}/H_D \quad (F)$$

The exact groups $H_D$ and $E_D$ are listed below.

| $D$ | $G$ | $H_D$ |
|---|---|---|
| 2 | $E_3 = SL(1,1) \otimes SL(2,1)$ | $SO(1,1) \otimes SO(2,1)$ |
| 3 | $E_4 = SL(2,3)$ | $SO(2,3)$ |
| 4 | $E_5 = SO(5,5)$ | $SO(5,\mathbb{C})$ |
| 5 | $E_6$ | $USp(8)$ |
| 6 | $E_7$ | $SU(8)$ |

Since these cosets describe gravity and its generalizations, time has also been "compactified". However, no dimensions have been dimensionally reduced or compactified in these theories because (1) the fields are still gauge fields (not scalars), and (2) sectioning implicitly eliminates winding modes. The group $G$ is then the generalization of linear coordinate transformations (gauge parameters linear in $x$), while $H$ is the generalization of the local Lorentz group.

As usual, there tends to be branching to smaller irreducible representations upon group reduction. (In the string description, which we won't further describe here, $S$ and $T$ use the worldsheet, while $M$ and $F$ use higher-dimensional worldvolumes.)

Of course, supergravity has also fermions, and there is no guarantee that all bosons will appear in $G/H$, even in the minimally supersymmetric case. For these purposes, we will also need to consider the related superspaces for all these theories. The anticommuting coordinates $\theta$ are irreducible spinors of $H$ (and of the internal R-symmetry).

Just as in general relativity, where a global linear coordinate transformation $GL(D)$ can be generalized to an arbitrary coordinate transformation, one generalizes a global transformation of $E_{D+1}$ to a "generalized coordinate" transformation. The generalized coordinate transformation acts on the vielbiens as [13, 14]

$$\delta_U E_A{}^N = U^M \partial_M E_A{}^N - \partial_M U^N E_A{}^M + \eta^{NP}{}_\mathfrak{m} \eta^\mathfrak{m}{}_{QR} \partial_P U^Q E_A{}^R.$$

The first two terms are just an ordinary coordinate transformation in general relativity, while in the last term $\eta$ is the structure constants of the current algebra. Its form depends on $D$, and the indices $\mathfrak{m}$ represent the indices for worldvolume coordinates. For $D = 3$ it is **5** in $SL(5)$ and for $D = 4$ it is **10** for $SO(5,5)$. See [7, 13] for a complete list of $\eta$ for various $D$. For the gauge transformation to be closed, we need to impose the section condition.

$$\eta^{NM}{}_\mathfrak{m} \partial_{1M} \partial_{2N} = 0.$$

The different subscripts just indicate that the derivative can act on different coordinates [13, 14].

## 2.2 4-pt Amplitudes

In 4D Yang-Mills theory, the color ordered 4-point tree amplitude can be written in a form that is manifestly gauge invariant with respect to linearized gauge invariance of the external fields [16]

$$A_4 = \frac{F^4}{st},$$

where $F$ is the linearized field strength

$$F_{mn} = \partial_{[m} A_{n]},$$

and the indices are contracted in a way that all double poles are canceled. In the corresponding 1-loop amplitude in the maximally supersymmetric case, the kinematic factors are the same, and the $1/st$ is replaced with the 1-loop scalar box diagram [16]. Similar remarks apply to higher loops for the leading order in $1/N$ [18, 19].

In terms of the spinor notation, it becomes

$$\frac{f^2 \bar{f}^2}{st},$$

where

$$f_{\alpha\beta} = \partial_{(\alpha}{}^{\dot{\gamma}} A_{\beta)\dot{\gamma}}$$

is the selfdual part of the YM field strength, while the anti-selfdual part $f_{\dot{\alpha}\dot{\beta}}$ can be defined similarly.

The 4-pt gravity tree level amplitude can be written in a similar form [16]:

$$\frac{W^4}{stu}.$$

$W$ is the linearized Weyl tensor $W_{[ab][cd]}$:

$$W^{[mn]}_{[pq]} = \partial_{[p} \partial^{[m} h^{n]}_{q]}.$$

Just like the YM 4-pt amplitude, it can also be written in its spinor form. The 4-pt amplitude in this form is

$$A_4(1, 2, 3, 4) = [(w_1 w_2)(\bar{w}_3 \bar{w}_4) + (\bar{w}_1 \bar{w}_2)(w_3 w_4) + (2 \leftrightarrow 3) + (2 \leftrightarrow 4)] \frac{1}{stu}$$

The Weyl tensor in spinor indices is

$$w_{(\alpha\beta\gamma\delta)} = \partial_{(\alpha}{}^{\dot{\varepsilon}} \partial_{\beta}{}^{\dot{\zeta}} h_{\gamma\delta)\dot{\varepsilon}\dot{\zeta}}, \quad \text{c.c.}$$

We have explicitly written down the contractions between the Weyl tensors since we are going to use this amplitude later. This amplitude contains all six possible configurations of helicities.

The 4-pt amplitude given above can be generalized to the 4-pt amplitude for type II superstrings:

$$A_{4\text{superstring}} = W^4 C(s, t, u),$$

where

$$C(s, t, u) = \pi \frac{\Gamma(-\alpha's/2)\Gamma(-\alpha't/2)\Gamma(-\alpha'u/2)}{\Gamma(1 + \alpha's/2)\Gamma(1 + \alpha't/2)\Gamma(1 + \alpha'u/2)}.$$

To find the 4-pt F-theory amplitude, we need first to find the analog of the Weyl tensor in F-theory, and find the correct contractions between the Weyl tensors such that (1) it is gauge invariant and (2) it reduces to the ordinary gravity amplitude after reduction.

The reason that the first condition is not trivial is that, as we will see, there is in general no gauge invariant analog of the Weyl tensor in F-theory, so we need to check the gauge invariance independently. With the help of the twistor formalism, one can find unique contractions that satisfy both conditions above.

## 3 Twistors

To analyze the spectrum of on-shell states, we will also need the little group $H$ of each $H$. For the usual Lorentz groups this can be found by reducing the number of space and time dimensions each by 1; for the covering groups (with the spinor as the defining representation), this is essentially reducing the range of the spinor index by half. For our theories, we just look at the known descriptions of each theory in 2 lower dimensions, and Wick rotate $H$ for these $D - 2$ dimensional theories to compact groups (for Abelian factors, phases):

$$H_D = H_{D-2,\text{compact}}$$

Covariant on-shell field strengths, since they have no gauge degrees of freedom, are easily related to independent physical degrees of freedom using twistors (or supertwistors). The twistors $\lambda$ are irreducible spinors of both $H$ and little $H$ (direct product). They are elements of the corresponding coset space, related to $H/H$. More precisely, they are elements of the related projective space, found by starting from the massless bispinor momentum in the lightcone frame, and looking at the rectangular piece of the spinor-representation $H$ group element needed to transform to arbitrary frames for both groups. (This twistor will be constrained for orthogonal or symplectic covering groups.) The on-shell momentum is thus quadratic in twistors (with a different twistor for each momentum). The twistors relate covariant field strengths to physical degrees of freedom by replacing $H$-spinor indices with $H$-spinor indices.

Note that reduction to the little group by use of twistors is equivalent to going to the lightcone gauge: In that gauge, the remaining components of the gauge field are the same as the independent components of the field strength (up to factors of $p^+$). Thus the field strength with little-group indices is equivalent to the "transverse" gauge field.

## 3.1 Dimensions

Explicitly in $D = 3$, all the relevant groups, as well as their anticommuting coordinates, twistors, and momenta quadratic in twistors (where ( ) means symmetrized), are [15, 20–22]

|   | $G$ | $H$ | $H$ | $R$ | $\theta$ | $\lambda$ | $p$ |
|---|---|---|---|---|---|---|---|
| F | $SL(5)$ | $O(3,2) = Sp(4)$ | $SO(2)$ | $I$ | $\boldsymbol{\alpha}$ | $a\boldsymbol{\alpha}$ | $(\boldsymbol{\alpha\beta})$ |
| T | $O(3,3) = SL(4)$ | $O(2,1)^2 = Sp(2)^2$ | $I$ | $I$ | $\alpha, \alpha'$ | $\alpha, \alpha'$ | $(\alpha\beta), (\alpha'\beta')$ |
| M | $GL(4)$ | $O(3,1) = Sp(2,C)$ | $U(1)$ | $U(1)$ | $\alpha, \dot\alpha$ | $\alpha, \dot\alpha$ | $\alpha\dot\beta$ |
| S | $GL(3)$ | $O(2,1) = Sp(2)$ | $I$ | $SO(2)$ | $a\alpha$ | $\alpha$ | $(\alpha\beta)$ |

(Of course, for 2×2 matrices $Sp = SL$.) We have written irreducible spinors of $H$ (defining representations of the covering group) with Greek indices, and those of $H$ (on $\lambda$) or R-symmetry (on $\theta$) with Latin. The boldface Greek indices mean four-component spinors, while the unbold Greek are two-component spinors.

(U(1) $H$ indices on $\lambda$ are single-valued and not displayed.)

Note that $Sp(4)$ is 4-component real, so the $SO(2)$ index is 2-component real, while $Sp(2,C)$ is 2-component complex, so the $U(1)$ index is implicitly 1-component complex. (However, we can use a complex basis for $SO(2)$, with off-diagonal metric.) Of course, the symmetric $SO(2)$ metric is used to construct $F$'s $p$ from $\lambda$'s.

For $D = 4$ we have (where $\langle \ \rangle$ means antisymmetrized $Sp$-traceless) [15, 20–22]

|   | $G$ | $H$ | $H$ | $R$ | $\theta$ | $\lambda$ | $p$ |
|---|---|---|---|---|---|---|---|
| F | $O(5,5)$ | $O(5,C) = Sp(4,C)$ | $O(3)O(2) = U(2)$ | $U(1)$ | $\boldsymbol{\alpha}, \dot{\boldsymbol{\alpha}}$ | $a\boldsymbol{\alpha}, a\dot{\boldsymbol{\alpha}}$ | $\boldsymbol{\alpha\dot\beta}$ |
| T | $O(4,4)$ | $O(3,1)^2 = Sp(2,C)^2$ | $O(2)^2 = U(1)^2$ | $U(1)^2$ | $\alpha, \dot\alpha, \alpha', \dot\alpha'$ | $\alpha, \dot\alpha, \alpha', \dot\alpha'$ | $\alpha\dot\beta, \alpha'\dot\beta'$ |
| M | $GL(5)$ | $O(4,1) = USp(2,2)$ | $O(3) = USp(2)$ | $USp(2)$ | $a\boldsymbol{\alpha}$ | $a\boldsymbol{\alpha}$ | $\langle\boldsymbol{\alpha\beta}\rangle$ |
| S | $GL(4)$ | $O(3,1) = Sp(2,C)$ | $O(2) = U(1)$ | $U(2)$ | $a\alpha, a\dot\alpha$ | $\alpha, \dot\alpha$ | $\alpha\dot\beta$ |

(and $USp(2) = SU(2)$).

The appearance of momenta in the supersymmetry algebras and expression in terms of twistors can be read from these tables, but are listed in Appendix B for convenience.

## 3.2 Symplectic constraint

There are always 4 $\theta$'s in $D = 3$ and 8 in $D = 4$, but the number of $\lambda$'s is

$$[\lambda] = 2^{\text{rank}(G)-1},$$

ranging from twice the number of $\theta$'s to half. (It decreases by a factor of 2 with each vertical step down in the diamond.) Note: All $\lambda$'s here are $2n \times n$ matrices, though there may be more than 1 such matrix.

Since all these twistors are symplectic with respect to $H$ (a characteristic of $D = 3$ and 4), they also satisfy the symplectic constraint (see Appendix C)

$$\lambda_a^\alpha \lambda_{b\alpha} = 0$$

(with indices contracted with the symplectic metric). Because of antisymmetry in $ab$, this is trivial for single-valued indices ($I$ and $U(1)$), and a single component for double-valued indices ($SO(2)$ and $Sp(2)$).

At least for $D = 3$ and 4, this constraint implies

$$p^2 = 0,$$

i.e., contracting a single pair of spinor indices between 2 $p$'s gives 0. Again for these dimensions, that means only $p^2 = 0$ for $S$ and $M$, and $p^2 = p'^2 = 0$ for $T$, essentially because all these $p$'s have $Sp(2)$ spinor indices. But the above constraint is stronger for $F$, whose $p$ has $Sp(4)$ indices.

The condition $p^2 = 0$ is actually a special case of the section condition in F-theory. The section condition in F-theory is necessary to keep the gauge transformation of F-theory closed. The section condition (in any dimension) is in the form $\partial \otimes \partial = 0$, where the two $\partial$ act on different functions. In momentum space, this means that the section condition involves the product of two different momenta. Thus the corresponding condition in terms of the twistors will also involve two different twistors.

In 3D F-theory, the section condition is

$$p_{1\langle\alpha}^\gamma p_{2\gamma\beta\rangle} = 0.$$

We shall rewrite this condition in terms of the twistors. For the convenience of calculation, we will use $U(1)$ indices for the little group instead of $SO(2)$, i.e. we will write $\lambda_\alpha \equiv \lambda_\alpha^1 + i\lambda_\alpha^2$, and $\bar\lambda_\alpha \equiv \lambda_\alpha^1 - i\lambda_\alpha^2$. The momentum is $p_{(\alpha\beta)} = \bar\lambda_{(\alpha}\lambda_{\beta)}$, and the section condition is

$$\frac{1}{4}\left(\bar\lambda_{1\gamma}\lambda_2^\gamma\bar\lambda_{1\langle\alpha}\lambda_{2\beta\rangle} + \bar\lambda_{1\gamma}\lambda_2^\gamma\lambda_{1\langle\alpha}\bar\lambda_{2\beta\rangle}\right) + \frac{1}{4}\left(\bar\lambda_{1\gamma}\bar\lambda_2^\gamma\lambda_{1\langle\alpha}\lambda_{2\beta\rangle} + \lambda_{1\gamma}\lambda_2^\gamma\bar\lambda_{1\langle\alpha}\bar\lambda_{2\beta\rangle}\right) = 0.$$

A consistent constraint on twistors is

$$\lambda_{1\gamma}\bar\lambda_2^\gamma = 0, \quad \lambda_{1\gamma}\lambda_2^\gamma\bar\lambda_{1\langle\alpha}\bar\lambda_{2\beta\rangle} + \text{c.c.} = 0. \tag{3.1}$$

By letting $\lambda_1 = \lambda_2$, we recover our original symplectic constraint.

As we will see in later sections, this particular choice of constraints is consistent when we reduce the twistors into their M and T-theory forms. We can also write the above constraint in terms of $SO(2)$ indices,

$$\lambda_{1\gamma}^{[a}\lambda_2^{\gamma b]} = 0, \quad \lambda_{1\gamma a}\lambda_2^{\gamma a} = 0, \quad \lambda_{1\gamma\{a}\lambda_{2b\}}^{\gamma}\lambda_{1\langle\alpha}^{\{a}\lambda_{2\beta\rangle}^{b\}} = 0,$$

where $\{ab\}$ is the symmetric and traceless part of $SO(2)$ indices.

## 3.3 On-shell momenta

On-shell momenta, and thus twistors after applying this constraint and gauge invariance, correspond to off-shell momenta in 1 less (time) dimension. We can check this by comparing counting for known off-shell momenta in one theory to our new counting for twistors of the same theory in one dimension higher.

First we compare the counting of off-shell 3D $p$'s to on-shell 4D $p$'s, with the latter defined by imposing gauge transformations and the symplectic constraint to the twistors:

|   | 3D $p$ | 4D $\lambda$ − gauge − constraint |
|---|---|---|
| F | 10 | $16 - 4 - 2$ |
| T | 6  | $8 - 2 - 0$ |
| M | 4  | $8 - 3 - 1$ |
| S | 3  | $4 - 1 - 0$ |

The group theory analysis for $D = 2$ is a bit simple, so we'll just do the counting comparison for off-shell 2D and on-shell 3D:

|   | 2D $p$ | 3D $\lambda$ − gauge − constraint |
|---|---|---|
| F | 6 | $8 - 1 - 1$ |
| T | 4 | $4 - 0 - 0$ |
| M | 3 | $4 - 1 - 0$ |
| S | 2 | $2 - 0 - 0$ |

# 4 Fields

## 4.1 General remarks

We will concentrate on the generalization $h(x)$ of the linearized metric, and have some comments on the supersymmetric fields. $h$ lives in the coset space $G/H$ as a Lie algebra, since it is the perturbation of vielbein under flat space: $E_A{}^M = \delta_A{}^M + h_A{}^M$. For on-shell physics, we have to break the symmetry to $H$, which means that we can lower the index $M$, and there is no distinction between the curved and flat indices. (We shall replace $A$ with $N$.) The antisymmetric part of $h_{[MN]}$ is in the adjoint representation of $H$ and thus it is gauged away. The remaining symmetric part is the representation of $h$. In $D = 3$, this is **14** in $Sp(4, \mathbb{R})$ (i.e. the $2 \times 2$ Young diagram). In spinor indices, it is represented as $h_{\langle\alpha\beta\rangle\langle\gamma\delta\rangle}$. For $D = 4$, the representation of $h$ is $\mathbf{5} \times \bar{\mathbf{5}}$; in spinor notation, it is $h_{\langle\alpha\beta\rangle\langle\dot\gamma\dot\delta\rangle}$. For the notation, see Appendix A. For S and M-theory, $h$ is the usual metric (in $D$ and $D+1$ dimensions, respectively), while for T-theory this is the direct product of left and right vectors (each in its own $D$ dimensions), i.e., the NS-NS sector. However, for $S$ and $M$, we'll take $G$ as $SL$ and not $GL$, since the determinant of the metric is an auxiliary field on shell, where all fields are essentially conformal.

The gauge parameter $\zeta$ has the same index structure as the momentum $p$, so the form of the (linearized) gauge transformation $\delta h = \partial \zeta$ is unique.

The physical states in any (super)field strength $\psi$ follow from converting the Greek spinor indices ($H$) to Latin spinor indices ($H$) of half the range, by use of the twistors. This means contracting either spinor index on $p$ with any index on $\psi$ kills it, by the symplectic constraint. Doing the same for $\theta$ replaces it with the fermionic partner of the twistor $\lambda$, which are always 1/2 in the number of that of $\theta$, but satisfy a Clifford algebra. This implies the universal massless equation $\displaystyle{\not}p q = 0$ for supersymmetry $q$ (implied by unitarity), as well as the same constraint $\displaystyle{\not}p d = 0$ for the covariant spinor derivative $d$. It also implies contracting the spinor index on $d$ with any index on $\psi$ kills it, so we have both conditions of the form

$$p^{\alpha\beta}\psi_{\alpha\ldots} = d^\alpha \psi_{\alpha\ldots} = 0$$

with various types of spinor indices.

Although the general consideration above works in any dimension, we shall focus on the 3-dimensional case for the remainder of the paper; in the derivation of the 4-pt amplitude, we will work directly with the Weyl tensor $w$, instead of the supersymmetric field strength $\psi$.

## 4.2 3D

In $D = 3$, $h$ always appears at order $\theta^2$ in a bispinor superfield prepotential $H(x, \theta)$. However $D = 3$, $N = 0, 1, 2$ (super)gravity contain no physical degrees of freedom. As a result, in S and T-theory the physical states appear in a vector multiplet described by a scalar prepotential $V$. On the other hand, in M and F-theory no scalar prepotential appears on shell.

This description automatically follows from the reduction from F-theory:

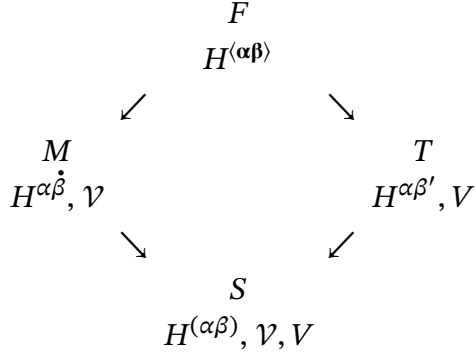

Thus there are always 5 prepotentials (and always 4 $\theta$'s), but only $H^{\langle\alpha\beta\rangle}, H^{\alpha\dot\beta}, V$ survive on shell, while $H^{\alpha\beta'}, H^{(\alpha\beta)}, \mathcal{V}$ do not. [1]

Consequently, in F and M-theory, the generalized Weyl tensor $w(x)$ appears at linear order in $\theta$ in a trispinor superfield strength $W(x,\theta)$, which at lowest order contains the conformal part of the gravitino field strength. We use these on-shell field strengths here only at the linearized level, for external states. For T and F they are not defined off shell, and we'll need to use the constraint $p^2 = 0$ to obtain gauge-invariant field strengths. But in T-theory the surviving vector multiplet is described by a bispinor superfield strength, where 1 "scalar" appears at lowest order in $\theta$ in a Maxwell-like tensor and the other at order $\theta^2$ in a Weyl-like tensor $w$. (Dimensional reduction generally produces field strengths of lower engineering dimension.) And in S-theory the superfield strength is a true scalar, with a Maxwell tensor for the other bosonic degree of freedom at order $\theta^2$. (There is no Weyl tensor in ordinary 3D gravity.)

The following table gives the index structure of these various fields, including the $\theta^2$ coefficient of $h$ in $H$ (and for convenience we have repeated $p$):

|   | $p$ | $h$ | $H$ | $\theta^2$ | $w$ | $W$ |
|---|---|---|---|---|---|---|
| F | $(\alpha\beta)$ | $\langle\alpha\beta\rangle\langle\gamma\delta\rangle$ | $\langle\alpha\beta\rangle$ | $\langle\gamma\delta\rangle$ | $(\alpha\beta\gamma\delta)$ | $(\alpha\beta\gamma)$ |
| T | $(\alpha\beta), (\alpha'\beta')$ | $(\alpha\beta)(\gamma'\delta')$ | $\alpha\gamma'$ | $\beta\delta'$ | $(\alpha\beta)(\gamma'\delta')$ | $\alpha\beta'$ |
| M | $\alpha\dot\beta$ | $(\alpha\beta)(\dot\gamma\dot\delta)$ | $\alpha\dot\gamma$ | $\beta\dot\delta$ | $(\alpha\beta\gamma\delta)$ ($\oplus$h.c.) | $(\alpha\beta\gamma)$ |
| S | $(\alpha\beta)$ | $(\alpha\beta\gamma\delta)$ | $(\alpha\beta)$ | $a(\gamma a\delta)$ | 0 | scalar |

(The $a$ pairs have been contracted with the $SL(2)$, not $O(2)$, metric of $SO(2)$. We use either $\langle\alpha\beta\rangle\langle\gamma\delta\rangle$ or $(\alpha\beta)(\gamma\delta)$ to refer to the 2×2 Young tableau of $Sp(4)$ of dimension 14, the traceless, symmetric, second-rank tensor of $O(5)$.) For T, the gravitino field strength has an index structure intermediate between those of $W$ and $w$. For S, the fermions are an ordinary spinor at linear order in $\theta$ in the scalar $W$, with independent degrees of freedom stopping at order $\theta^2$.

## 4.3 4D

In 4D, the generalized metric $h(x)$ appears at order $\theta^4$ in a scalar superfield prepotential $V(x,\theta)$. Then the generalized Weyl tensor $w(x)$ appears at order $\theta^2$ in a bispinor superfield strength $W(x,\theta)$, the gravitino field strength appears at linear order in $\theta$, and some "spin-1" field strength appears at $\theta = 0$.

The index structure of these fields, as well as the $\theta^2$ coefficient of $w$ in $W$, are:

|   | $p$ | $h$ | $w$ | $W$ | $\theta^2$ |
|---|---|---|---|---|---|
| F | $\boldsymbol{\alpha\dot\beta}$ | $\langle\boldsymbol{\alpha\beta}\rangle\langle\boldsymbol{\dot\gamma\dot\delta}\rangle$ | $(\boldsymbol{\alpha\gamma})(\boldsymbol{\dot\beta\dot\delta})$ | $\boldsymbol{\alpha\dot\beta}$ | $\boldsymbol{\gamma\dot\delta}$ |
|   |   |   | $\langle\alpha\beta\rangle\langle\gamma\delta\rangle\,(\oplus\text{h.c.})$ | $\langle\alpha\beta\rangle$ | $\langle\gamma\delta\rangle$ |
| T | $\alpha\dot\beta, \alpha'\dot\beta'$ | $\alpha\dot\beta\gamma'\dot\delta'$ | $(\alpha\gamma)(\dot\beta'\dot\delta')\,(\oplus\text{h.c.})$ | $\alpha\dot\beta'$ | $\gamma\dot\delta'$ |
|   |   |   | $(\alpha\gamma)(\beta'\delta')\,(\oplus\text{h.c.})$ | $\alpha\beta'$ | $\gamma\delta'$ |
| M | $\langle\boldsymbol{\alpha\beta}\rangle$ | $(\boldsymbol{\alpha\beta})(\boldsymbol{\gamma\delta})$ | $(\boldsymbol{\alpha\beta\gamma\delta})$ | $(\boldsymbol{\alpha\beta})$ | $a(\boldsymbol{\gamma}a\boldsymbol{\delta})$ |
| S | $\alpha\dot\beta$ | $(\alpha\beta)(\dot\gamma\dot\delta)$ | $(\alpha\beta\gamma\delta)\,(\oplus\text{h.c.})$ | $(\alpha\beta)$ | $a(\gamma a\delta)$ |

Since $V$ is a scalar, its $\theta^4$ coefficient of $h$ has the same index structure as $h$ (and contractions of 2 pairs of Sp(2) indices in the last 2 cases).

(Note: In replacing indices $\alpha \to a$ to find physical states, $\langle\boldsymbol{\alpha\beta}\rangle \to [ab]$, using the symplectic constraint, where [ ] means antisymmetrized.)

## 4.4 On shell field strengths

The field strength for the metric (Weyl tensor) can be written in the form $w = \partial^2 h$. The precise index structure can be deduced by requiring the correct representation of $w$. Usually, there will only be one or two possible index structures that can match.

In 3D F-theory $w_{(\boldsymbol{\alpha\beta\gamma\delta})}$, written in terms of $h$, can only be a linear combination of the following index structure:

$$\partial_\alpha{}^\epsilon \partial_\gamma{}^\zeta h_{<\beta\epsilon><\delta\zeta>} + (\text{sym. perm. over } \boldsymbol{\alpha\beta\gamma\delta}), \text{ and}$$

$$\partial_{\alpha\gamma} \partial^{\epsilon\zeta} h_{<\beta\epsilon><\delta\zeta>} + (\text{sym. perm. over } \boldsymbol{\alpha\beta\gamma\delta}).$$

Unfortunately, no linear combination of the above two index structures can achieve gauge invariance. This seems to be consistent with the fact that no gauge invariant tensor in exceptional geometry analogous to the Riemann curvature tensor is found [12, 14].

While a gauge invariant Weyl tensor cannot be constructed, we shall see that expressions such as $w_1 w_2$ or $w_1 w_2 w_3 w_4$ are gauge invariant, as long as all the indices are contracted. (Different subscripts implies that the momenta are different in these Weyl tensors.)

To represent the on-shell Weyl tensor in terms of twistors, we need to write the metric $h$ in terms of the twistor first. In 3D M-theory (that is, ordinary 4D gravity), the on-shell metric is expressed in polarization vectors

$$h_{(\alpha\beta)(\dot\gamma\dot\delta)} \propto \varepsilon^+_{\alpha\dot\gamma}\varepsilon^+_{\beta\dot\delta} + \varepsilon^-_{\alpha\dot\gamma}\varepsilon^-_{\beta\dot\delta} \ .$$

The polarization vectors must satisfy the transverse and traceless conditions,

$$p \cdot \varepsilon^\pm = 0, \quad (\varepsilon^\pm)^2 = 0.$$

If we choose the polarization vectors to be in the form

$$\varepsilon^+_{\alpha\dot\beta} = \lambda_\alpha \bar\eta_{\dot\beta}, \quad \varepsilon^-_{\alpha\dot\beta} = \bar\lambda_{\dot\beta} \chi_\alpha,$$

then they satisfy the traceless and transverse conditions. $\eta, \chi$ are reference twistors that contain the degrees of freedom of 4D gravity.

Note that the polarization vectors have their own residual gauge transformation

$$\varepsilon^\pm \to \varepsilon^\pm + c_\pm p.$$

This translates into freedom in choosing $\chi, \eta$

$$\chi_\alpha \to \chi_\alpha + c_- \lambda_\alpha. \quad \bar\eta_{\dot\beta} \to \bar\eta_{\dot\beta} + c_+ \bar\lambda_{\dot\beta}.$$

Therefore, the total degrees of freedom in $\eta$ and $\chi$ are $2 + 2 - 1 - 1 = 2$, which is exactly the same as $h$; $h$ can be expressed as

$$h_{(\alpha\beta)(\dot\gamma\dot\delta)} = \lambda_\alpha \lambda_\beta \bar\eta_{\dot\gamma} \bar\eta_{\dot\delta} + \chi_\alpha \chi_\beta \bar\lambda_{\dot\gamma} \bar\lambda_{\dot\delta}.$$

For convenience, we have chosen a normalization $\chi\lambda = 1 = \eta\lambda$. This makes $\eta, \chi$ have the dimension $m^{-1/2}$, and fixes the amplitude of wavefunctions of incoming and outgoing particles.

We wish to find a similar expression for the F-theory metric. The F-theory metric should contain the same on-shell degrees of freedom as the M-theory metric. When we reduce the metric from F to M-theory, only the $h_{(\alpha\beta)(\dot\gamma\dot\delta)}$ component should be non-zero, and equals the expression we just derived. The correct answer is

$$h_{\langle\alpha\gamma\rangle\langle\beta\delta\rangle} = \left(\lambda_{\langle\alpha}\bar\eta_{\gamma\rangle}\lambda_{\langle\beta}\bar\eta_{\delta\rangle} + \bar\lambda_{\langle\alpha}\chi_{\gamma\rangle}\bar\lambda_{\langle\beta}\chi_{\delta\rangle}\right) + (\boldsymbol{\alpha} \leftrightarrow \boldsymbol{\beta}) + (\boldsymbol{\gamma} \leftrightarrow \boldsymbol{\delta}) \quad (4.1)$$

As we will see in the next section, it will reduce to the correct form in M-theory. Just like the case of 4D gravity, these reference twistors have residual gauge transformations $\eta \to \eta + c\lambda$, and $\chi \to \chi + c\lambda$.

Plug the result into the two possible index structures of $w$. For the first index structure, we get

$$\partial^\varepsilon_{(\alpha} \partial^\zeta_\beta h_{\gamma\delta)\varepsilon\zeta} = \lambda_{(\alpha}\lambda_\beta \lambda_\gamma \lambda_{\delta)} + \bar\lambda_{(\alpha}\bar\lambda_\beta \bar\lambda_\gamma \bar\lambda_{\delta)}.$$

For the second index structure, we get

$$\partial_{(\alpha\beta}\partial^{\varepsilon\zeta}h_{\gamma\delta)\varepsilon\zeta} = 0.$$

Therefore, the second index structure will not contribute to the Weyl tensor on-shell:

$$w_{(\alpha\beta\gamma\delta)} = \partial^{\varepsilon}_{(\alpha}\partial^{\zeta}_{\beta}h_{\gamma\delta)\varepsilon\zeta} = \lambda_{(\alpha}\lambda_{\beta}\lambda_{\gamma}\lambda_{\delta)} + \bar{\lambda}_{(\alpha}\bar{\lambda}_{\beta}\bar{\lambda}_{\gamma}\bar{\lambda}_{\delta)}.$$

The final form of the Weyl tensor does not depend on the reference twistors.

# 5 Reduction

## 5.1 From F to M

Reducing from F-theory to M-theory is breaking $H$ from $Sp(4,\mathbb{R})$ to $SO(3,1)$. In spinor indices, it is equivalent to breaking a symplectic 4-component spinor $\boldsymbol{\alpha}$ into two 2-component $SL(2,\mathbb{C})$ spinors $\alpha, \dot{\alpha}$. The twistor will break down accordingly:

$$\lambda_{\boldsymbol{\alpha}}, \bar{\lambda}_{\boldsymbol{\alpha}} \quad \to \quad \bar{\lambda}_{\alpha}, \bar{\lambda}_{\dot{\alpha}}, \lambda_{\alpha}, \lambda_{\dot{\alpha}}.$$

In terms of the partial derivative (and therefore the spacetime coordinates) it would be reduced as

$$\partial_{(\boldsymbol{\alpha\beta})} \to \partial_{\alpha\dot{\beta}}, \partial_{(\dot{\alpha}\dot{\beta})}, \partial_{(\alpha\beta)}.$$

$\partial_{(\dot{\alpha}\dot{\beta})}$ and $\partial_{(\alpha\beta)}$ can be set to zero by the section condition $p^{\gamma}_{1\langle\alpha}p_{2\gamma\beta\rangle} = 0$. Observe that they can also be set to zero by applying the solution

$$\bar{\lambda}_{\alpha} = \lambda_{\dot{\alpha}} = 0,$$

and the only remaining twistors are $\lambda_{\alpha}, \bar{\lambda}_{\dot{\alpha}}$, which are the ordinary 4D gravity twistors.

Since there is no section condition in 3D M-theory (4D gravity), the above solution should entirely solve the symplectic constraint; we find that this is true.

For the first condition in (3.1), we see that

$$\lambda_{\boldsymbol{\alpha}}\bar{\eta}^{\boldsymbol{\alpha}} \quad \to \quad \lambda_{\alpha}\bar{\eta}^{\alpha} + \lambda_{\dot{\alpha}}\bar{\eta}^{\dot{\alpha}} = 0.$$

For the second condition, we first break down $\lambda_{\langle\boldsymbol{\alpha}}\eta_{\boldsymbol{\beta}\rangle}$. Note that **5** in $Sp(4)$ breaks into $\mathbf{2 \times \bar{2} + 1}$:

$$\lambda_{\langle\boldsymbol{\alpha}}\eta_{\boldsymbol{\beta}\rangle} \quad \to \quad \lambda_{[\dot{\alpha}}\eta_{\beta]} + \frac{1}{4}\varepsilon_{\alpha\beta}(\lambda_{\gamma}\eta^{\gamma} - \bar{\lambda}_{\dot{\gamma}}\bar{\eta}^{\dot{\gamma}}) = \frac{1}{4}\varepsilon_{\alpha\beta}\lambda_{\alpha}\eta^{\alpha}.$$

Similarly, we find
$$\bar{\lambda}_{\langle\alpha}\bar{\eta}_{\beta\rangle} \quad \to \quad -\frac{1}{4}\varepsilon_{\alpha\beta}\bar{\lambda}_{\dot{\gamma}}\bar{\eta}^{\dot{\gamma}}$$

Thus one can see that the constraint for twistors indeed solves the section condition:
$$\lambda_{\langle\alpha}\eta_{\beta\rangle}\bar{\lambda}_{\gamma}\bar{\eta}^{\gamma} + \text{c.c} \quad \to \quad \frac{1}{4}\bar{\lambda}_{\dot{\alpha}}\bar{\eta}^{\dot{\alpha}}\lambda_{\alpha}\eta^{\alpha} - \frac{1}{4}\lambda_{\alpha}\eta^{\alpha}\bar{\lambda}_{\dot{\alpha}}\bar{\eta}^{\dot{\alpha}} = 0.$$

After reduction, the metric decomposes into $3 \times \bar{3} + 2 \times \bar{2} + 1$:
$$\underset{14}{h_{\langle\alpha\beta\rangle\langle\gamma\delta\rangle}} \quad \to \quad \underset{3\times\bar{3}}{h_{(\alpha\gamma)(\dot{\beta}\dot{\delta})}} \quad \underset{2\times\bar{2}}{h_{\alpha\dot{\beta}}} \quad \underset{1}{h.}$$

$h_{(\alpha\gamma)(\dot{\beta}\dot{\delta})}$ is the ordinary 4D metric (3D M-theory is just 4D supergravity), and $h$ the dilaton. $h_{(\alpha\dot{\beta})}$ is a 3-form tensor multiplet. (In 4D a 3-form is dual to a vector.) The fact that it is a 3-form can be seen from the gauge transformation of $h_{(\alpha\dot{\beta})}$.

Only the $3 \times \bar{3}$ part should be non-zero on shell, and it should just be the ordinary 4D metric. We find that this is indeed the case. This can seen by using the fact that $\lambda_{\langle\alpha}\bar{\eta}_{\beta\rangle} \to \lambda_{\alpha}\eta_{\dot{\beta}}$ under reduction; (4.1) will indeed become ordinary 4D matrices.

Decomposing the Weyl tensor:
$$\underset{35}{w_{(\alpha\beta\gamma\delta)}} \quad \to \quad \underset{5+\bar{5}}{w_{(\alpha\beta\gamma\delta)}, w_{(\dot{\alpha}\dot{\beta}\dot{\gamma}\dot{\delta})}} \quad \underset{2\times\bar{4}+4\times\bar{2}}{w_{\dot{\alpha}(\beta\gamma\delta)}, w_{\alpha(\dot{\beta}\dot{\gamma}\dot{\delta})}} \quad \underset{3\times\bar{3}}{w_{(\alpha\beta)(\dot{\gamma}\dot{\delta})}}.$$

$w_{(\alpha\beta\gamma\delta)}$ and $w_{(\dot{\alpha}\dot{\beta}\dot{\gamma}\dot{\delta})}$ are the usual Weyl tensor for the metric, and $w_{\dot{\alpha}(\beta\gamma\delta)}, w_{\alpha(\dot{\beta}\dot{\gamma}\dot{\delta})}$ are the field strengths for the 3-form, and $w_{(\alpha\beta)(\dot{\gamma}\dot{\delta})}$ is the field strength for $h$.

Both $2 \times \bar{4} + 4 \times \bar{2}$ and $3 \times \bar{3}$ part will become zero on shell, and the $5 + \bar{5}$ part will survive.
$$w_{(\alpha\beta\gamma\delta)} = \lambda_{(\alpha}\lambda_{\beta}\lambda_{\gamma}\lambda_{\delta)}, \quad w_{(\dot{\alpha}\dot{\beta}\dot{\gamma}\dot{\delta})} = \bar{\lambda}_{(\dot{\alpha}}\bar{\lambda}_{\dot{\beta}}\bar{\lambda}_{\dot{\gamma}}\bar{\lambda}_{\dot{\delta})}$$

It is the Weyl tensor for 4D gravity, which is as expected.

## 5.2 From F to T

Reducing from F-theory to T-theory is breaking the isotropy group from $Sp(4, \mathbb{R})$ to $SO(2, 2) = SO(3) \times SO(3)$. Group theoretically, it is identical to the $F \to T$ case. The only difference is $F \to M$ breaks down one time coordinate and $F \to T$ breaks down one space coordinate. In spinor indices, a symplectic 4-component spinor $\alpha$ is reduced into two 2-component $SO(3)$ spinors $\alpha, \alpha'$. The twistor becomes:
$$\lambda_{\alpha}, \bar{\lambda}_{\alpha} \quad \to \lambda_{\alpha}, \lambda_{\alpha'}, \bar{\lambda}_{\alpha}, \bar{\lambda}_{\alpha'}$$

The partial derivative decomposes into

$$\partial_{(\alpha\beta)} \to \partial_{\alpha\beta'}, \partial_{(\alpha'\beta')}, \partial_{(\alpha\beta)}.$$

By the section condition, $\partial_{\alpha\beta'}$ is zero, while $\partial_{(\alpha'\beta')}, \partial_{(\alpha\beta)}$ remain. Just as the $F \to M$ case, one can find some solution in terms of the twistors which will bring $\partial_{\alpha\beta'}$ to zero. Such a solution is

$$\bar{\lambda}_\alpha = \lambda_\alpha, \quad \bar{\lambda}_{\alpha'} = -\lambda_{\alpha'}. \tag{5.1}$$

The only remaining independent twistors are $\lambda_\alpha$ and $\lambda_{\alpha'}$.

Note that since T-theory is not section condition free, this solution will not completely solve the section condition; but we will show that the original F-theory section condition will reduce to the T-theory section condition after applying the above solution. In terms of momenta, the T-theory section condition is

$$p_{1\alpha\beta} p_2^{\alpha\beta} + p_{1\alpha'\beta'} p_2^{\alpha'\beta'} = 0,$$

or in terms of the twistors

$$(\lambda_{1\alpha} \lambda_2^\alpha + \lambda_{1\alpha'} \lambda_2^{\alpha'})(\lambda_{1\alpha} \lambda_2^\alpha - \lambda_{1\alpha'} \lambda_2^{\alpha'}) = 0.$$

The first F-theory twistor constraint in (3.1) is solved by (5.1):

$$\lambda_{\boldsymbol{\alpha}} \bar{\eta}^{\boldsymbol{\alpha}} \to \lambda_\alpha \bar{\eta}^\alpha + \lambda_{\alpha'} \bar{\eta}'^\alpha = 0.$$

For the second constraint, it will break from **5** into $(\mathbf{2, 2}) + (\mathbf{1, 1})$. The $(\mathbf{2, 2})$ part is also completely solved:

$$\lambda_{[\alpha} \eta_{\beta']}(\bar{\lambda}_\alpha \bar{\eta}^\alpha - \bar{\lambda}_{\alpha'} \bar{\eta}^{\alpha'}) + \text{c.c} \to 0.$$

On the other hand the $(\mathbf{1, 1})$ part will not vanish:

$$(\lambda_\alpha \eta^\alpha + \lambda_{\alpha'} \eta^{\alpha'})(\bar{\lambda}_\alpha \bar{\eta}^\alpha - \bar{\lambda}_{\alpha'} \bar{\eta}^{\alpha'}) + c.c \to 2(\lambda_\alpha \eta^\alpha + \lambda_{\alpha'} \eta^{\alpha'})(\lambda_\alpha \eta^\alpha - \lambda_{\alpha'} \eta^{\alpha'}),$$

but it is exactly the section condition for T-theory.

Similarly, we can reduce the F-theory metric and Weyl tensor to its T-theory counterpart:

$$\underset{14}{h_{<\boldsymbol{\alpha\beta}><\boldsymbol{\gamma\delta}>}} \to \underset{(\mathbf{3,3'})}{h_{(\alpha\gamma)(\beta'\delta')}} \quad \underset{(\mathbf{2,2'})}{h_{(\alpha\beta')}} \quad \underset{\mathbf{1}}{h.}$$

and the Weyl tensor becomes

$$\underset{35}{w_{(\boldsymbol{\alpha\beta\gamma\delta})}} \to \underset{\mathbf{5+5'}}{w_{(\alpha\beta\gamma\delta)}, w_{(\alpha'\beta'\gamma'\delta')}} \quad \underset{\mathbf{2\times 4' + 4\times 2'}}{w_{\alpha'(\beta\gamma\delta)}, w_{\alpha(\beta'\gamma'\delta')}} \quad \underset{\mathbf{3\times 3'}}{w_{(\alpha\beta)(\gamma\delta')}}.$$

# 6 4-point S-matrices

## 6.1 Gauge invariance

As we mentioned in the previous section, the Weyl tensor itself is not gauge invariant, even on shell and under the section condition. Under the gauge transformation of $h$

$$\delta h_{\langle\alpha\beta\rangle\langle\gamma\delta\rangle} = \partial^{\langle\gamma}_{\langle\alpha}\xi^{\delta\rangle}_{\beta\rangle},$$

the variance of the Weyl tensor $w$ is complicated; by using the section condition, one can see that $\delta w$ must be in the form:

$$\delta w_{(\alpha\beta\gamma\delta)} = c_1 \partial_{(\alpha\beta}\partial_{\gamma\delta)}\partial^{\epsilon\zeta}\xi_{\epsilon\zeta} + c_2 \partial_{(\alpha\beta}\partial^\epsilon_\gamma \partial^\zeta_{\delta)}\xi_{\epsilon\zeta}.$$

This is because any contraction between the partial derivatives vanishes. Therefore, the above form exhausts all the possible contractions of $\partial^3 \xi$ that are compatible with the symmetries.

If we write the partial derivative in terms of twistors, we get

$$\delta w_{(\alpha\beta\gamma\delta)} = c_1 \lambda_{(\alpha}\lambda_\beta \bar\lambda_\gamma \bar\lambda_{\delta)}(\lambda_\epsilon \lambda_\zeta \xi^{\epsilon\zeta}) + c_2 \left[\lambda_{(\alpha}\lambda_\beta \lambda_\gamma \bar\lambda_{\delta)}(\bar\lambda_\epsilon \lambda_\zeta \xi^{\epsilon\zeta}) + \text{c.c}\right].$$

Note that both terms in $\delta w$ have at least one uncontracted $\lambda$ and one uncontracted $\bar\lambda$. This will be important later for proving gauge invariance.

Now we wish to prove that $w_{1\alpha\beta\gamma\delta}w_2^{\alpha\beta\gamma\delta}$ is gauge invariant. First, observe that $\delta(w_1 w_2) = \delta w_1 w_2 + w_1 \delta w_2$, and $w = \lambda^4 + \bar\lambda^4$. Using the fact that $\delta w$ contains at least one uncontracted $\lambda$ and one uncontracted $\bar\lambda$, there must be either a $\lambda_1 \bar\lambda_2$ or $\lambda_2 \bar\lambda_1$ in every term of $\delta w_1 w_2$. Therefore it vanishes by the section condition. The same argument applies for $w_1 \delta w_2$.

This argument can apply to any contractions $w_1 w_2 \cdots w_n$, as long as all indices are contracted. Without loss of generality, we assume that $w_1 w_2 \cdots w_n$ cannot be factored into smaller pieces, since any index contractions of the form $w_1 w_2 \cdots w_n$ can be decomposed into products of terms that cannot be decomposed further.

The gauge transformation is

$$\delta(w_1 w_2 \cdots w_n) = (\delta w_1) w_2 \cdots w_n + \cdots$$

Note that since $w_2 \cdots w_n$ cannot be decomposed into smaller pieces, it can only contain terms with all $\lambda$ or terms with all $\bar\lambda$; any terms with both $\lambda$ and $\bar\lambda$ will disappear by the section condition. And since $\delta w_1$ contains both uncontracted $\lambda$ and $\bar\lambda$, every term will vanish. By this argument, we can see that $w_1 w_2 \cdots w_n$ is indeed gauge invariant.

## 6.2 Amplitudes

To find the correct amplitude, we just have to find the correct combinations of $w^4$ that will reduce to the ordinary 4-pt graviton amplitude when going from F to M-theory.

The 4-pt graviton (superstring) amplitude is

$$A_4(1,2,3,4) = [(w_1 w_2)\bar{w}_3 \bar{w}_4 + \bar{w}_1 \bar{w}_2 w_3 w_4 + (2 \leftrightarrow 3) + (2 \leftrightarrow 4)] C(s,t,u),$$

where

$$C(s,t,u) = \pi \frac{\Gamma(-\alpha' s/2)\Gamma(-\alpha' t/2)\Gamma(-\alpha' u/2)}{\Gamma(1+\alpha' s/2)\Gamma(1+\alpha' t/2)\Gamma(1+\alpha' u/2)}.$$

Using $w_{\alpha\beta\gamma\delta} = \lambda_{(\alpha}\lambda_\beta\lambda_\gamma\lambda_{\delta)}$ and $\bar{w}_{\dot\alpha\dot\beta\dot\gamma\dot\delta} = \lambda_{(\dot\alpha}\lambda_{\dot\beta}\lambda_{\dot\gamma}\lambda_{\dot\delta)}$, the amplitude will decompose to the sum of all possible helicity combinations. (Only 2 positive and 2 negative helicities have non-zero amplitudes; that's the reason why we have the $w^2 \bar{w}^2$ structure.) For notational simplicity, we will use the following notation for 4D twistors: $\lambda_{1\alpha} = \langle 1|, \lambda_1^\alpha = |1\rangle$, and $\bar{\lambda}_{\dot\alpha} = |1], \bar{\lambda}_1^{\dot\alpha} = [1|$.

$s, t, u$ here are the Mandelstam variables. There are some useful relations between Mandelstam variables and twistors:

$$\begin{aligned} s &= -(p_1+p_2)^2 = -2\langle 12\rangle[12] \\ t &= -(p_1+p_3)^2 = -2\langle 13\rangle[13] \\ u &= -(p_1+p_4)^2 = -2\langle 14\rangle[14] \end{aligned}$$

We observe that when reducing from F to M-theory, the Weyl tensor $\mathbf{w}$ in F-theory reduces to $\mathbf{w} \to \bar{w} + w$. (In this section, we use boldface $\mathbf{w}$ to denote the Weyl tensor of F-theory and non-boldface $w$ to denote the Weyl tensor in M theory.) Therefore the only way to get a $\bar{w}^2 w^2$ term is to have a $(\mathbf{w})^2(\mathbf{w})^2$ index structure in F-theory; any other contractions of $\mathbf{w}^4$ will give $\bar{w}^4 + w^4$.

The contraction $(\mathbf{w})^2(\mathbf{w})^2$ will also generate $w^4 + \bar{w}^4$ terms upon reduction, which is unwanted. To remedy this, we have to put additional $\mathbf{w}^4$ terms that will cancel the $\bar{w}^4 + w^4$ terms after reduction.

To be more specific, we first consider the following candidate for F-theory amplitude:

$$A_4 = [(\mathbf{w}_1 \mathbf{w}_2)(\mathbf{w}_3 \mathbf{w}_4) + (2 \leftrightarrow 3) + (2 \leftrightarrow 4)] C(s,t,u)$$

After reduction, it becomes

$$A_4 = [w_1 w_2 \bar{w}_3 \bar{w}_4 + (w_1 w_2)(w_3 w_4)] C(s,t,u) + \text{c.c.} + (2 \leftrightarrow 3) + (2 \leftrightarrow 4).$$

In terms of twistor notation, it becomes

$$A_4 = \left[\langle 12\rangle^4[34]^4 + \langle 34\rangle^4[12]^4 + \langle 12\rangle^4\langle 34\rangle^4 + [12]^4[34]^4\right] C(s,t,u) + (2 \leftrightarrow 3) + (2 \leftrightarrow 4).$$

There are additional unwanted parts

$$([12]^4[34]^4 + \text{c.c.})C(s,t,u) + (2 \leftrightarrow 3) + (2 \leftrightarrow 4).$$

We can simplify it by using the relation

$$\frac{\langle 13\rangle\langle 24\rangle}{\langle 12\rangle\langle 34\rangle} = \frac{\langle 13\rangle[13]\langle 24\rangle[24]}{\langle 12\rangle[13]\langle 34\rangle[24]} = -\frac{t}{s}.$$

This can be seen using the relation between twistors and Mandelstam variables given above and momentum conservation:

$$\langle 12\rangle[13] = -\langle 24\rangle[43].$$

Similarly, we have,

$$\frac{\langle 14\rangle\langle 32\rangle}{\langle 12\rangle\langle 34\rangle} = -\frac{u}{s},$$

and their complex conjugate counterparts. The unwanted part can be simplified to

$$\frac{[12]^4[34]^4 + \text{c.c.}}{stu}\left(\frac{s^4 + t^4 + u^4}{s^4}\right).$$

Using $s + t + u = 0$, we get the following relation

$$s^4 + t^4 + u^4 = -2(s^2t^2 + t^2u^2 + u^2s^2).$$

It can be verified that the following contraction of $\mathbf{w}^4$ will lead to this exact structure when reducing to M-theory:

$$w_{1\alpha\beta\gamma\delta}w_2^{\gamma\delta\varepsilon\zeta}w_{3\varepsilon\zeta\rho\pi}w_4^{\rho\pi\alpha\beta}.$$

We can see this by direct calculation very similar to the one we did above:

$$\mathbf{w}^4 \underset{\text{reduction}}{\Rightarrow} (\langle 12\rangle^2\langle 34\rangle^2\langle 13\rangle^2\langle 24\rangle^2 + \text{c.c})C(s,t,u) + (2 \leftrightarrow 3) + (2 \leftrightarrow 4)$$

$$= (\langle 12\rangle^4\langle 34\rangle^4 + [12]^4[34]^4)C(s,t,u)\left(\frac{s^2t^2 + t^2u^2 + u^2s^2}{s^4}\right).$$

The full form of the 3D F-theory 4-pt amplitude with all the indices written out is

$$\left[(w_{1\alpha\beta\gamma\delta}w_2^{\alpha\beta\gamma\delta})(w_{3\varepsilon\zeta\rho\pi}w_4^{\varepsilon\zeta\rho\pi}) - 2w_{1\alpha\beta\gamma\delta}w_2^{\gamma\delta\varepsilon\zeta}w_{3\varepsilon\zeta\rho\pi}w_4^{\rho\pi\alpha\beta}\right]C(s,t,u) + (2 \leftrightarrow 3) + (2 \leftrightarrow 4).$$

In terms of twistors, it is

$$\left[(\lambda_1\lambda_2)^4(\lambda_3\lambda_4)^4 - 2(\lambda_1\lambda_2)^2(\lambda_2\lambda_4)^2(\lambda_4\lambda_3)^2(\lambda_3\lambda_1)^2 + \text{c.c}\right]C(s,t,u) + (2 \leftrightarrow 3) + (2 \leftrightarrow 4).$$

# 7 Conclusions

In this paper, we have developed the twistor formalism for F-theory and applied it to find the 4-pt tree amplitude of F-theory. The crucial part of finding the amplitude is to find the correct contractions between the 4 Weyl tensors that will be both gauge invariant and reduce to the 4-pt graviton amplitude under the reduction $F \to M$. The twistor formalism helped us make the derivation simple.

There are two directions one can generalize the result. One is to work directly in the superspace formalism given in [2], where the superamplitude is directly calculated. Another direction is to consider higher dimensions, where the four-point amplitude is not just the pure graviton amplitude, but a compact form consisting of all the 4-pt interactions of scalars, vectors, and form fields.

To find the superamplitude, one has to consider the F-theory version of the super-Weyl tensor $W_{\alpha\beta\gamma}$. Just as the same case with the ordinary Weyl tensor, one can not generally find the gauge invariant form of $W$, which is confirmed in [2]. Note that since $W$ has dimension 5/2, it should be related to the (on-shell) prepotential by $W = d^5 H$. One has to enumerate all the possible combinations of $d^5 H$, up to the section condition and Bianchi Identity. It also has to satisfy the following: (1) It has to become the ordinary super-Weyl tensor under the reduction to M-theory. (2) All the scalar contractions of $W$'s have to be gauge invariant. (3) It should relate to our bosonic Weyl tensor $w_{\alpha\beta\gamma\delta} = d_{(\alpha} W_{\beta\gamma\delta)}$. The kinetic factor of the final superamplitude is just $\int d^4\theta W^4$.

To find the higher dimensional F-theory amplitude, one needs a better way to find the F-theory analog of the Weyl tensor. Since the index structure is much more complicated in higher dimensions, one cannot rely on the bootstrap method to determine the unique Weyl tensor. One possible approach is to solve the torsion constraint of the current algebra (see [8] for the first step in this direction), but this is expected to be computationally complicated.

# Acknowledgments

WS is supported by NSF grant PHY-1620628.

# A  Conventions

Two-component spinor indices are denoted by Greek letters $\alpha, \beta, \gamma$... ($SL(2, \mathbb{C})$ indices). Four-component indices are denoted using boldface Greek letters $\boldsymbol{\alpha}, \boldsymbol{\beta}, \boldsymbol{\gamma}$... ($Sp(4)$ or $Sp(4, \mathbb{C})$ indices).

Vector indices are denoted by Latin letters. For two-component vectors ($SO(2)$ or $SL(2)$), we use $a, b, c$.., while for 4-component vectors ($SO(4)$ etc.) we use $m, n, p$..., and boldface Latin letters for 5-component vector indices **m, n**, .... The complex conjugate of a representation will be denoted by adding a dot ( ˙ ) on the indices. If there are two indices that are the same representation but transform independently, then we will add a prime ( ′ ) to distinguish them.

For the vector indices of F-theory for general dimension $D$, we will use capital Latin letters: $A, B, C..$ for flat indices and $M, N...$ for curved indices. Their representations are listed below:

| $D$ | Representation (in $H$) |
|---|---|
| 2 | $(\mathbf{2}, \mathbf{3})$ i.e. $a\alpha$ |
| 3 | $\mathbf{10}$ i.e. $\langle\alpha\beta\rangle$ |
| 4 | $\mathbf{4} + \bar{\mathbf{4}}$ i.e. $\alpha\dot{\beta}$ |
| 5 | $\mathbf{27}$ |
| 6 | $\mathbf{28} + \bar{\mathbf{28}}$ |

The convention for the symmetry of the indices are as follows:

$$\text{Symmetric: } (ab) \tag{A.1}$$
$$\text{Antisymmetric: } [\alpha\beta] \tag{A.2}$$
$$\text{Traceless Symmetric: } \{ab\} \tag{A.3}$$
$$\text{Traceless Antisymmetric: } \langle\alpha\beta\rangle \tag{A.4}$$

# B  Supersymmetry algebras and on-shell momenta

For completeness we list the supersymmetry algebras for $D = 3$:

$$F: \quad \{q_\alpha, q_\beta\} = p_{(\alpha\beta)} \tag{B.1}$$
$$T: \quad \{q_\alpha, q_\beta\} = p_{(\alpha\beta)}, \quad \{q_{\alpha'}, q_{\beta'}\} = p_{(\alpha'\beta')} \tag{B.2}$$
$$M: \quad \{q_\alpha, \bar{q}_{\dot\beta}\} = p_{\alpha\dot\beta} \tag{B.3}$$
$$S: \quad \{q_{a\alpha}, q_{b\beta}\} = \delta_{ab} p_{(\alpha\beta)} \tag{B.4}$$

and for $D = 4$:

$$F: \quad \{q_\alpha, \bar{q}_{\dot\beta}\} = p_{\alpha\dot\beta} \tag{B.5}$$
$$T: \quad \{q_\alpha, \bar{q}_{\dot\beta}\} = p_{\alpha\dot\beta}, \quad \{q_{\alpha'}, \bar{q}_{\dot\beta'}\} = p_{\alpha'\dot\beta'} \tag{B.6}$$
$$M: \quad \{q_{a\alpha}, q_{b\beta}\} = C_{ab} p_{\langle\alpha\beta\rangle} \tag{B.7}$$
$$S: \quad \{q_{a\alpha}, \bar{q}^b_{\dot\beta}\} = \delta^b_a p_{\alpha\dot\beta} \tag{B.8}$$

where again all *C*'s are symplectic (Hermitian) metrics.

We also have similar expressions for the on-shell momenta in terms of twistors (except R-symmetry is replaced with *H*): again for $D = 3$,

$$F: \quad p_{(\alpha\beta)} = \delta^{ab}\lambda_{a\alpha}\lambda_{b\beta} \quad (C^{\beta\alpha}\lambda_{a\alpha}\lambda_{b\beta} = 0) \tag{B.9}$$

$$T: \quad p_{(\alpha\beta)} = \lambda_\alpha\lambda_\beta, \quad p_{(\alpha'\beta')} = \lambda_{\alpha'}\lambda_{\beta'} \tag{B.10}$$

$$M: \quad p_{\alpha\dot\beta} = \lambda_\alpha\lambda_{\dot\beta} \tag{B.11}$$

$$S: \quad p_{(\alpha\beta)} = \lambda_\alpha\lambda_\beta \tag{B.12}$$

and 4:

$$F: \quad p_{\alpha\dot\beta} = \lambda_{a\alpha}\bar\lambda^a_{\dot\beta} \quad (C^{\beta\alpha}\lambda_{a\alpha}\lambda_{b\beta} = \bar C^{\dot\beta\dot\alpha}\bar\lambda^a_{\dot\alpha}\bar\lambda^b_{\dot\beta} = 0) \tag{B.13}$$

$$T: \quad p_{\alpha\dot\beta} = \lambda_\alpha\bar\lambda_{\dot\beta}, \quad p_{\alpha'\dot\beta'} = \lambda_{\alpha'}\bar\lambda_{\dot\beta'} \tag{B.14}$$

$$M: \quad p_{\langle\alpha\beta\rangle} = C^{ba}\lambda_{a\alpha}\lambda_{b\beta} \quad (C^{\beta\alpha}\lambda_{a\alpha}\lambda_{b\beta} = 0) \tag{B.15}$$

$$S: \quad p_{\alpha\dot\beta} = \lambda_\alpha\bar\lambda_{\dot\beta} \tag{B.16}$$

# C  Symplectic projective spaces

The twistors we describe here use the following construction; similar methods apply to orthogonal and linear groups. (We skip the discussion of reality properties.) Symplectic group elements $g$ satisfy

$$g^T C g = C, \quad C^T = -C$$

in terms of the symplectic metric $C$ ($C^2 = I$). Dividing $g$ into 2 equal rectangles by a vertical axis

$$g = (\lambda, \kappa)$$

we choose a corresponding basis for the metric

$$C = i\begin{pmatrix} 0 & I \\ -I & 0 \end{pmatrix}$$

Then we find in particular

$$\lambda^T C \lambda = 0$$

The $GL(n)$ subgroup of $Sp(2n)$ is given by

$$g \to \begin{pmatrix} M & 0 \\ 0 & M^{T-1} \end{pmatrix}$$

for arbitrary matrix $M$, due to the block-off-diagonal form chosen for $C$. (A block-diagonal $C$ for $n$ even would lead naturally instead to an $Sp(n)^2$ subgroup.) Choosing $g$ to transform from the left by the global group $G$ ($Sp(2n)$) and from the right by the local group $H$ ($GL(n)$), the twistor $\lambda$ transforms as

$$\lambda' = g_0 \lambda h$$

For $n$ even we can then directly also choose $H$ to be the $Sp(n)$ subgroup of $GL(n)$ instead.

These constructions can be phrased completely in the language of coset spaces by enlarging $H$ to gauge away $\kappa$.